\documentclass[fleqn,twoside]{article}
\usepackage{espcrc2}

\newcommand{\bdm}{\begin{displaymath}}
\newcommand{\edm}{\end{displaymath}}
\newcommand{\no}{\nonumber \\}

\renewcommand{\Re}{{\rm Re}\,}

\newcommand{\be}{\begin{equation}}
\newcommand{\ee}{\end{equation}}
\newcommand{\bea}{\begin{eqnarray}}
\newcommand{\eea}{\end{eqnarray}}
\newcommand{\fs}{\; \; .}
\newcommand{\co}{\; \; ,}

\newcommand{\la}{\langle}
\newcommand{\ra}{\rangle}

\newcommand{\lbar}{\bar{\ell}}

\newcommand{\epe}{\varepsilon'/\varepsilon}

\usepackage{graphicx}
\usepackage[figuresright]{rotating}

\newcommand{\AmS}{{\protect\the\textfont2
  A\kern-.1667em\lower.5ex\hbox{M}\kern-.125emS}}

\title{Chiral perturbation theory,
dispersion relations and final state interactions
in $K \to \pi \pi$}

\author{Gilberto Colangelo\address{Institut f\"ur Theoretische Physik
    der Universit\"at Z\"urich\\ Winterthurerstr. 190, 8057 Z\"urich,
    Switzerland}}
       
\begin{document}

\begin{abstract}
We discuss the recent literature on the treatment of final state
interactions in $K \to \pi \pi$. Various approaches are compared and
particular emphasis is given to the possibility of combining dispersive
methods with lattice input. Recent results on the dependence of various
quantities on the quark masses at order $p^6$ in the chiral expansion are
presented and the relevance for the lattice calculations is discussed.
\end{abstract}

\maketitle

\section{INTRODUCTION}
Ideally, lattice QCD calculations should provide numbers that can be
directly compared to the experimental ones. In practice, one often needs
extra theoretical input and/or treatment before being able to compare to
the phenomenology. One important example of this is the $K \to \pi \pi$
amplitude: until the recent work by Lellouch and L\"uscher \cite{lellu} it
was not even known how one could directly calculate this amplitude on the
lattice, and the problems of principle related to a naive approach had been
identified by Maiani and Testa in a famous no-go theorem \cite{MT}. The
standard method to work around this no-go theorem consists in calculating
on the lattice an unphysical matrix element, $K \to \pi$, and to use an
approximate relation (based on tree-level chiral perturbation theory
(CHPT)) to obtain the $K \to \pi \pi$ amplitude \cite{bernard}. The latter
step is known to represent a rather crude approximation, and sizeable
higher order CHPT corrections are expected. Alternative methods to obtain
the $K \to \pi \pi$ amplitude were discussed in \cite{rome}.  Despite its
theoretical cleanliness, the use of the Lellouch--L\"uscher method
represents a formidable challenge, and is still out of reach for today's
state-of-the art numerical simulations. Indeed all lattice collaborations
\cite{lattice_cn} that have presented results for the $K \to \pi \pi$
amplitude at this conference have used either the method of Bernard et
al. \cite{bernard} or one of those proposed in \cite{rome}.

In view of this, it is worth to ask whether one could devise a better way
to obtain the $K \to \pi \pi$ amplitude from quantities that are calculable
today on the lattice. For example, since today's calculations are all done
in the quenched approximation, one may try to find a way to obtain
information about the unquenched results from (partially) quenched
calculations. Work in this direction has already been done, see
e.g. Ref. \cite{quenched}, but I will not touch this issue here.  Another
very important issue is that of final state interactions (FSI): if one
calculates the $K \to \pi$ matrix element and obtains the $K \to \pi \pi$
amplitude by using tree level CHPT, one is neglecting FSI. Although these
unitarity effects are formally of higher order in the chiral counting, they
are known to be particularly large when two pions interact in the $S$ wave
in the isospin zero channel, and might be crucial in the final numerical
result. The role of FSI has been discussed at length in the recent
literature, particularly in connection to the calculations of $\epe$ in the
Standard Model \cite{PP,PPS,Buras,kpp,offshell}. The emphasis of the recent
literature is on the dispersive treatment of the FSI, but all the
groups that calculated $\epe$ making explicit use of the chiral expansion
beyond leading order, did also find large corrections due to FSI
\cite{trieste,hambye}. Earlier treatments of $K \to \pi \pi$, and in
particular of the $CP$ conserving part of the amplitude also discussed the
importance of FSI, and its role for the $\Delta I=1/2$ rule
\cite{truong,kambor}.

In this talk I will review the recent and the less recent literature on the
treatment of FSI in $K \to \pi \pi$. I will do so by comparing two
different approaches: one which explicitly relies on CHPT, and one that
uses only dispersion relations. A third possibility, which is also
discussed, consists in a dispersion-relation inspired improvement of a
chiral calculation. The focus of this discussion will be on the methods,
rather than on the numerical results for the $K \to \pi \pi$ amplitude,
which I will not touch at all. 
A second aspect that I will discuss in this talk is
the dependence of some quantities on the quark masses, as it is evaluated
in CHPT at next-to-next-to-leading order. I will present the analyses which
are currently available \cite{pipi,ABT}, for masses, decay constants and
scattering amplitudes. The latter result is directly connected to the
problem of FSI: it has been shown that large FSI are also reflected into a
very strong dependence on the quark masses \cite{pipi}.

\section{THE SCALAR FORM FACTOR: CHPT vs. DISPERSIVE METHODS}
In order to compare the chiral expansion and the dispersive approach, it is
easier to consider a simpler quantity than the $K \to \pi \pi$ amplitude --
if we want two pions in the $S$ wave and isospin zero in the final state,
the simplest quantity we can think of is the scalar form factor of the
pion: 
\be
\Gamma(s) = {\cal N} \langle \pi(p_1) \pi(p_2)| \bar u u + \bar d d|
0 \rangle \co
\ee
where $s=(p_1+p_2)^2$ and ${\cal N}$ ensures $\Gamma(0)=1$.
In CHPT this has been calculated (and studied numerically) to two loops
\cite{FF}, and has also been analyzed with dispersive methods \cite{DGL},
so that we can immediately compare the two results. We first consider the
CHPT calculation, which reads, to one loop \cite{GL84}
\be
\Gamma(s)\!=\!1+{s \over N} (\lbar_4\! -\!1)
+{1 \over 2 F_\pi^2} (2s-M_\pi^2 ) \bar J(s) +O(s^2)\; ,
\ee
where $N=16 \pi^2 F_\pi^2$, and $\bar J(s)$ is a loop function \cite{GL84}.
In order to compare to the dispersive treatment, it is useful to rewrite
this expression in a different way, that makes the $\pi \pi$ phase shift
appear explicitly
\be
\Gamma(s)=1+c s + {s^2 \over \pi} \int_{4 M_\pi^2}^\infty ds'
{\delta^{(2)}(s') \over s'^2 (s'-s)} +O(s^2) \; ,
\label{eq:fschdr}
\ee
where $c$ is a constant related to $\lbar_4 \sim \ln M_\pi^2$, and where
$\delta^{(2)}$ is the leading order $\pi \pi$ phase shift in the $S$ wave
and $I=0$ channel
\bea
\delta(s)&=&\delta^{(2)}(s)+\delta^{(4)}(s)+\ldots \no
\delta^{(2)}(s)&=&{\pi \sigma(s) \over 2 N} (2 s - M_\pi^2) \co
\label{eq:delta2}
\eea
where $\sigma(s)=\sqrt{1- 4M_\pi^2/s}$.
\begin{figure}[tbh]
\begin{center}
\leavevmode
\includegraphics[width=7cm]{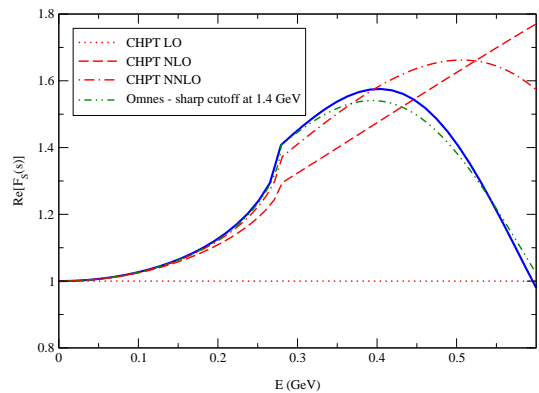}
\caption{\label{fig:sff} Scalar form factor: comparison of the CHPT
  calculation to leading, next-to-leading and next-to-next-to-leading order
  and the dispersive one (see legenda). The solid curve is the ``true form 
  factor'', see text.}  
\end{center}
\end{figure}
In order to have a numerical representation in CHPT to one loop, one needs
to fix one constant, $\lbar_4$. Once this is done, e.g. by fixing the
derivative of $F(s)$ at $s=0$ at its physical value (see below), the form
factor is as shown by the dashed line in Fig.~\ref{fig:sff}. The same kind
of analysis can be repeated to two loops -- there, however, one has to fix
more constants \cite{FF}. In Fig.~\ref{fig:sff} it is clearly seen that
while the chiral expansion behaves nicely up to around $s = (0.4~
\mbox{GeV})^2$, at higher energies it does not work so well. In particular,
the behaviour of the series at $s=M_K^2$ looks as follows:
\be
1~(\mbox{LO})~ \rightarrow ~1.62~(\mbox{NLO}) ~\rightarrow~
1.67~(\mbox{NNLO}) \co 
\ee
whereas the ``true value'' is $1.42$.
At slightly higher values, the two loop form factor does bend down, as the
true form factor does -- however, at a quantitative level it does not give
a satisfactory representation. It is not difficult to understand why two
loops are not yet enough to reproduce correctly the form factor at these
energies. The real part of the form factor is proportional to the cosine of
the phase
\be
\Re \Gamma(s) \sim \cos \delta(s) \sim 1 - {{\delta(s)}^2 \over 2} +
\ldots \co
\ee
and since the phase is a quantity of order $p^2$ in the chiral expansion,
this negative contribution to the real part starts only at
NNLO. Numerically, at NNLO the cosine becomes different from one, but it is
still larger than the real value by about 20 \% -- which roughly explains
the difference seen in Fig.~\ref{fig:sff}.

The dispersive treatment for the form factor is quite simple: Omn\`es
\cite{omnes} tells us that if one knows the phase $\phi$ of the form factor
(on the positive real axis, from $s=4M_\pi^2$ up to infinity), one knows
the form factor in the whole complex plane. Its explicit expression reads
\be
\Gamma(s)=\exp {s \over \pi} \int_{4 M_\pi^2}^\infty ds' {\phi(s') \over
  s'(s'-s)} \fs
\label{eq:om}
\ee
In principle one should worry about the presence of zeros, and consider a
corresponding polynomial factor, but for the sake of simplicity we will
disregard this possibility. We also assume that the integral in
Eq.~(\ref{eq:om}) converges.

Note that what enters in the Omn\`es function is the phase of the form
factor -- according to Watson's theorem, below the inelastic threshold this
is equal to the $\pi \pi$ phase shift in the $S$ wave and isospin
zero. Above that, however, it cannot be easily related to observable
quantities. Note also that although the inelastic threshold is formally at
$s=16 M_\pi^2$, the 4 $\pi$ channel is very weak, and can as well be
neglected. Around 1 GeV the much stronger 2 $K$ channel opens up, and
one cannot neglect inelastic effects anymore.

In fact what we have indicated until now as ``the true form factor'' is
nothing but the result of a dispersive treatment which explicitly takes
into account the contribution of the 2 $K$ channel, in a fully consistent
manner (coupled channel analysis), and using the available data
\cite{DGL}. In that case, one cannot express the form factor as simply as
in Eq.~(\ref{eq:om}). However, even neglecting the inelastic channels, and
using the $\pi \pi$ phase shift, e.g. as given in \cite{pipi}, all the way
up to a cut-off, one gets a decent description of the form factor, as shown
in Fig.~\ref{fig:sff}, where the cut off Omn\`es function
\be
\Omega(s)=\exp {s \over \pi} \int_{4 M_\pi^2}^{\Lambda^2} ds' {\delta(s')
  \over s'(s'-s)} \co
\ee
for $\Lambda=1.4$ GeV is plotted. The precise value of the cut-off does
matter for the final result, but not too much, as long as one is not taking
it too low.

It is interesting to compare at the algebraic level CHPT to one loop and
the Omn\`es representation: in order to do so we have to expand the latter
in a chiral series -- this is done by expanding the phase of the form
factor, and correspondingly, the exponential. If we do so, we end up
exactly with Eq.~(\ref{eq:fschdr}). Note that if we simply insert the
leading order chiral phase, Eq.~(\ref{eq:delta2}), in the dispersive
integral (\ref{eq:om}) we get a divergent result, and therefore are forced
to subtract it at least once more, as in Eq.~(\ref{eq:fschdr}). Note also
that the the dispersive representation explains also the presence of a
chiral logarithm in the constant $c$ in Eq.~(\ref{eq:fschdr}): with the
leading order phase the once-subtracted dispersive integral is infrared
divergent and explodes logarithmically as the pion mass goes to zero.

In summary, in the case of the scalar isoscalar form factor at energies
around $s=M_K^2$ it is clear that the chiral expansion is less economical,
and gives a poorer description than the dispersive treatment, even if the
latter is done neglecting inelastic effects. This motivates us to apply a
dispersive treatment also to the $K \to \pi \pi$ amplitude.

\section{THE $K \to \pi \pi$ AMPLITUDE}
It is easy to see an analogy between the $K \to \pi \pi$ amplitude and the
scalar form factor discussed in the previous section, as far as FSI are
concerned. On the other hand, the physical $K \to \pi \pi$ amplitude does
not depend on any kinematical variable, and the dispersive treatment of the
scalar form factor cannot be immediately applied to this amplitude: first
we have to extend the definition of the $K \to \pi \pi$ amplitude to make
it depend on one (or more) complex variables. There is no unique way
to do so, and we will discuss two different options. The chiral
calculation, however, does not need this extension, and the corresponding
treatment of FSI can be discussed directly for the physical amplitude.

\subsection{CHPT treatment of FSI}
In CHPT to one loop the $K \to \pi \pi$ amplitude reads as follows
\cite{kambor}
\bea
{\cal A} &=& {\cal A}^{(2)} \left\{ 1 + {1 \over 2 F_\pi^2}(2M_K^2-M_\pi^2)
   \bar J(M_K^2) \right. \no
 &&\left.-  {2M_K^2-M_\pi^2 \over 2N }  \left( \ln {M_\pi^2
         \over \mu^2} + 1 \right)  + \ldots \right\} \; ,
\label{eq:ACHPT}
\eea where ${\cal A}^{(2)}$ is the leading order amplitude, and where we
have written explicitly only the once subtracted loop function generated by
the two-pion exchange, and the accompanying chiral log. The ellipsis stands
for other loop functions and counterterms. The problem with this amplitude
is that in order to use it for numerical calculations one has to input
values for the weak (and strong) counterterms, which are largely unknown,
especially if one wants to identify the contributions of individual
operators in the weak Hamiltonian. Also, the identification of the
contribution of a certain loop diagram is ambiguous, not only because of
the presence of the scale $\mu$, but also because the break down into
individual loop diagrams depends on the parametrization chosen for the
fields in the effective lagrangian. Note, e.g., that logs coming from other
loop diagrams have to cancel the term proportional to $M_K^2 \ln M_\pi^2$ in
(\ref{eq:ACHPT}), so that the full amplitude has a well defined $m_u=m_d
\to 0$ limit. Despite these ambiguities, one may find instructive to
evaluate the pion loop contribution in (\ref{eq:ACHPT}): for $\mu \sim
M_\rho$ this contribution amounts to about 40 \%, which is indeed quite
large. On this basis, it has often been stated that in CHPT the main
one-loop correction to the $K \to \pi \pi$ amplitude comes from $\pi \pi$
FSI \cite{kambor,trieste,hambye}.  We stress that, because of the
ambiguities mentioned above, this statement cannot be made very precise.

Note the analogy to the case of the scalar form factor: indeed
Eq.~(\ref{eq:ACHPT}) looks completely analogous to Eq.~(\ref{eq:fschdr}),
for $s=M_K^2$. However, while in that case we could fix the counterterm and
the accompanying log (both contained in $\lbar_4$), with the value of the
scalar radius of the pion, here we do not have a corresponding quantity to
use as input. Numerical statements about this amplitude depend
in a crucial way on a theoretical estimate of the low energy constants
involved.

\subsection{Dispersive treatment: $K$ off-shell} 
In view of the analogy between Eq.~(\ref{eq:ACHPT}) and
Eq.~(\ref{eq:fschdr}) for $s=M_K^2$, it is tempting to make the momentum
squared of the kaon the variable on which to construct the dispersion
relation. This idea has been proposed by Truong \cite{truong}, and recently
revived by Pallante and Pich \cite{PP}. Although the idea is seductively
simple, it implies a number of serious problems, all related to the fact
that the off-shell extrapolations of a physical hadronic amplitude are
arbitrary, in contrast to on-shell amplitudes, which are unambiguous and
well defined. A discussion of the effects of this arbitrariness can be
found in Ref.~\cite{offshell}, and the interested reader is referred to
that paper for further details. Both papers \cite{truong,PP} ignore this
arbitrariness, and choose an unphysical off-shell extrapolation for the
kaon. In fact both papers quote for the ``off-shell amplitude'' at tree
level in CHPT 
\be 
\mbox{``} {\cal A}^{(2)}(s) = A(s-M_\pi^2)\mbox{"} \co
\label{eq:offshell}
\ee where $A$ is a constant. Such an extrapolation corresponds to using the
kaon field in the effective lagrangian for going off-shell. This choice is
unphysical: one could add a term that vanishes at the equations of motion
(and hence does not affect any physical observable), and correspondingly
change the off-shell amplitude (\ref{eq:offshell}) with terms proportional
to $s-M_K^2$. There is no way to prefer one choice over the other as long
as one is using the fields in the effective lagrangian: it is well known
that these do not have any physical significance and are simply integration
variables in the path integral \cite{GL84}. The calculation of the one-loop
corrections to (\ref{eq:offshell}), as done in \cite{PP} makes the sickness
of this extrapolation even more evident: a standard basis of the $O(p^4)$
counterterms, as e.g. given in \cite{EKW}, does not suffice to reabsorb the
divergences of the one loop diagrams. Off-shell extrapolations can be
unambiguously defined by using quark operators, such as the axial current
or the pseudoscalar density -- in that case, the off-shell amplitude to
leading order is different from (\ref{eq:offshell}), and does not change if
one adds to the lagrangian terms that vanish at the equations of motion
\cite{offshell}. One-loop corrections can also be unambiguously
calculated. 

In the approach of Refs. \cite{truong,PP} introducing the off-shell
extrapolation is unnecessary, and, as discussed above, misleading. At tree
level, their procedure amounts to multiplying the on-shell tree level
amplitude by the Omn\`es function evaluated at $s=M_K^2$:
${\cal A}^{(2)} \to {\cal A}^{(2)} \Omega(M_K^2)$.
At one loop one can extend this procedure in the following manner
\cite{PPS}: 
\be 
{\cal A}^{(2)+(4)} \to \left[ {\cal A}^{(2)+(4)}  -
  \Delta^{(2)}(M_K^2) \right] \Omega(M_K^2) \co 
\label{eq:PPS}
\ee
where $\Delta^{(2)}(M_K^2)$ is the $O(p^2)$ part of the Omn\`es function
\be
\Delta^{(2)}(s)= {s \over \pi} \int_{4M_\pi^2}^{\Lambda^2} ds'
{\delta^{(2)}(s') \over s'(s'-s)} \co
\label{eq:Delta}
\ee
which needs to be subtracted from the one-loop amplitude to avoid double
counting. Note that since the $O(p^2)$ phase shift grows linearly with $s$
(\ref{eq:delta2}), the dispersive integral (\ref{eq:Delta}) diverges when
$\Lambda \to \infty$ -- the final result depends on the choice of the
cut-off. 

While the effect due to FSI is quite substantial if compared to the tree
level chiral amplitude, the exponentiation of the FSI part of the one loop
correction gives a much less dramatic increase. If we use a cut-off
$\Lambda=1$ (1.3) GeV, we get $\Delta^{(2)}(M_K^2)=0.3$ (0.4), and
$\Omega(M_K^2)=1.4$, (1.65). The choice of the cut-off gives an uncertainty
which is comparable to the increase due to the exponentiation. Moreover, to
obtain a numerical estimate from such a calculation, one still needs an
estimate for the $O(p^4)$ counterterms \cite{PPS}, as in pure CHPT. The
necessity to refer to the chiral expansion and to its many constants is, in
my opinion, the main disadvantage of this method. While it is economical
and gives a quick estimate of the effect of FSI if one refers to tree level
CHPT, it becomes much more involved if one extends it to one loop.
In addition, at one loop there remain large uncertainties.

\subsection{Dispersive treatment: momentum-carrying Hamiltonian}
The necessity to refer to CHPT disappears if one chooses a different
extrapolation of the physical amplitude, and allows the weak Hamiltonian to
carry momentum, as has been suggested in \cite{kpp}. The method goes as
follows: consider the amplitude
\be 
_{I=0}\la \pi(p_1) \pi(p_2) | {\cal H}^{1/2}_W(0)|K(q_1) \ra =: T^+(s,t,u)
\label{eq:Adef}
\ee 
with the Mandelstam variables 
$s=(p_1\!+\!p_2)^2, \;t=(q_1\!-\!p_1)^2, \; u=(q_1\!-\!p_2)^2$,
related by $s\!+\!t\!+\!u=2M_\pi^2\!+\!M_K^2\!+\!q_2^2$, where $q_2$ is the
momentum carried by the weak Hamiltonian. The physical decay amplitude is
obtained by setting $q_2^\mu=0$ ($s=M_K^2$, $t=u=M_\pi^2$).

The dispersive treatment of a function of three complex variable is rather
complicated. However, the problem simplifies considerably if one neglects
the contribution of the imaginary parts of $D$ and higher waves. With this
approximation, the amplitude can be decomposed into a combination of
functions of a single variable:
\bea
T^+(s,t,u)&=&M_0(s)+\left\{ {1 \over 3}\left[N_0(t)+2 R_0(t)\right]
\right. \no
&+& \left. {1 \over 2}\left[ \left(\!s-u-{M_\pi^2 \Delta \over t} \!\right)
  N_1(t)\right] \right\} \no
&+& \Big\{ (t \leftrightarrow u) \Big\}
\label{eq:T+dec}
\eea
where $\Delta = M_K^2-M_\pi^2$.

In this approximation, the dispersion relation for the full amplitude is
transformed into a set of coupled dispersion relations of functions of a
single variables. These can be solved numerically, as shown in
\cite{kpp}. An important point concerns the overall number of subtraction
constants which is two for the amplitude $T^+$. In order to solve the
dispersion relations and calculate the $K \to \pi \pi$ amplitude, one has
to provide as input the two subtraction constants.  One of the two can be
given in terms of the $K \to \pi$ amplitude -- a soft--pion theorem relates
the amplitude at the soft pion point ($s=u=M_\pi^2$, $t=M_K^2$) to the $K
\to \pi$ amplitude up to terms of order $M_\pi^2$:
\be
-{{\cal A}(K \to \pi)\over 2 F_\pi}= a + 
{ \bar N \over 3} +{\cal O}(M_\pi^2) \; ,
\label{eq:soft pion point}
\ee 
where $\bar N=N_0(M_K^2)+2 R_0(M_K^2)$.  Notice that although the
process involves a kaon, the relation is based on the use of the $SU(2)$
symmetry, and therefore suffers from corrections of order $M_\pi^2$
only. The existence of this soft-pion theorem is a crucial advantage of
this approach, because it simplifies the calculation of one of the two
subtraction constants.  The key of the problem is how to calculate
$b$. This constant is related, e.g., to the derivative in $s$ of the
amplitude $T^+$ at the soft pion point. In \cite{kpp} a Ward identity that
relates this derivative to a $K \to \pi$ matrix element was derived. The
calculation of this matrix element would determine $b$. Alternatively, one
could try to obtain information on the $K \to \pi \pi$ amplitude at another
unphysical point (i.e. with meson masses at their physical values but with
the weak Hamiltonian carrying momentum) and use this input to constrain the
second subtraction constant.

\begin{figure}[t] 
\leavevmode \begin{center}
\includegraphics[width=7cm]{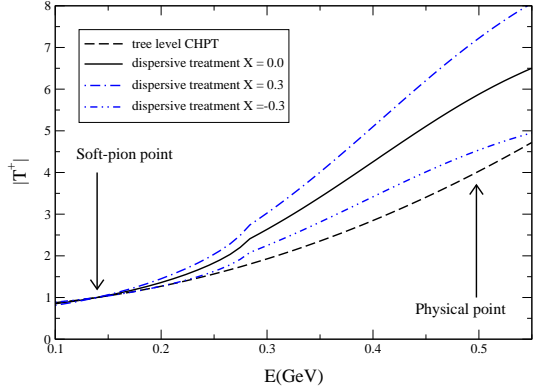}
\caption{\label{fig:kpp}The function $|T^+(s,t,u)|$ plotted {\em vs.}
  $E=\sqrt{s}$  along the line of constant $u=M_\pi^2$: the result of our
  numerical study for different values of $X$ are compared to tree level
  CHPT.}  
\end{center}
\end{figure}
In the absence of calculations of
$b$, the numerical results were illustrated by fixing $b$ at a certain value
and then varying it within a fairly wide range. To fix the central value
leading order CHPT was used: 
\be b= {3 a\over M_K^2-M_\pi^2} \left(1+ X + {\cal
O}(M_K^4) \right) \fs
\label{eq:ab_CHPT}
\ee 
The size of the correction $X$ is at the moment unknown, but nothing
protects it from being of order $M_K^2$.  In the numerical analysis the
generous range $X = \pm 30\%$ has been used. The results are shown in
Fig. \ref{fig:kpp}, where $|T^+(s,M_K^2+M_\pi^2-s,M_\pi^2)|$ versus $s$ is
plotted, comparing the numerical solution of the dispersion relations to
the CHPT leading order formula.  The latter is what has been used so far
whenever a number for the $K \to \pi \pi$ matrix element extracted from the
lattice has been given. The figure shows that large corrections with
respect to leading--order CHPT are to be expected. One source of large
corrections is the Omn\`es factor due to $\pi \pi$ rescattering in the
final state. The other potentially dangerous source is represented by $X$,
the next-to-leading order correction to the relation (\ref{eq:ab_CHPT})
between $a$ and $b$. The latter could (depending on the sign) in principle
double, or to a large extent reabsorb the correction due to final state
interactions.

It is clear that in order to obtain a calculation of $K \to \pi \pi$ with
this method one needs to determine these two constants -- and the necessary
information is not yet available. On the other hand, it is important to
emphasize the advantages of this method with respect to CHPT, or to the
Omn\`es improved version of it: here one does not need to introduce a large
number of constants, but only needs two inputs. The relation between these
constants and the $K \to \pi \pi$ amplitude is exact and it is provided by
the solution of the dispersion relation. The accuracy of the final result
will depend on the accuracy with which the two constants are known and the
accuracy in the solution of the dispersion relation. As far as the latter
is concerned the prospects of reducing it are quite good: the $\pi \pi$
phase shifts, which are dominating \cite{kpp}, are known with a remarkable
accuracy \cite{pipi}.  The numerical analysis performed so far \cite{kpp}
will be improved by taking into account the effects of the inelastic
contributions in the $s$ channel \cite{winp}.

\section{CHPT RESULTS ON QUARK MASS DEPENDENCE}
In CHPT the dependence on the quark masses is explicit: the low energy
constants of the effective lagrangian are quark-mass
independent. Therefore, once these are known, one can vary the quark masses
and see how any given quantity depends on them. This exercise is
particularly interesting from a lattice perspective, since the physical up
and down quark masses are still far from the reach of today's calculations,
and one may wish to learn beforehand what to expect, when those values are
approached. To make this exercise worthwhile it is important that one goes
beyond next-to-leading order: only in this manner can one check how fast
the series is converging and where one should stop believing the result of
the chiral expansion.  In the last few years there have been a few detailed
numerical analyses at next-to-next-to-leading order both in the
$SU(2)\times SU(2)$ and in the $SU(3) \times SU(3)$ framework, and I will
briefly review the results concerning the quark mass dependence.

\subsection{Chiral $SU(2) \times SU(2)$}
\begin{figure}[thb] 
\leavevmode \begin{center}
\includegraphics[width=7cm]{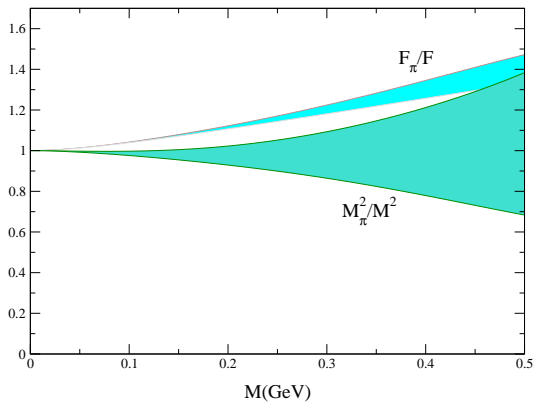}
\caption{\label{fig:fpi} Dependence of $F_\pi$ and $M_\pi$ on the average
  up and down quark masses $M=2B\hat m$.}
\end{center}
\end{figure}

In Ref.~\cite{pipi} the low energy constants appearing in the $\pi \pi$
scattering amplitude were determined by comparing the chiral representation
thereof to a phenomenological one that follows from Roy equations. The
numerical solution of these \cite{ACGL} provided the necessary input for
the determination. It has to be stressed that in this procedure, only the
constants that govern the momentum dependence of the amplitude could be
determined -- those that give the quark mass dependence had to be
fixed with theoretical estimates, and as such are subject to larger
uncertainties. 
\begin{figure}[thb] 
\leavevmode \begin{center}
\includegraphics[width=7cm]{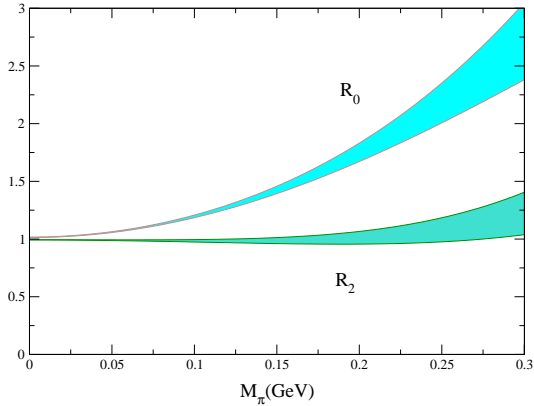}
\end{center}
\caption{\label{fig:a0}Dependence of the two $S$ wave scattering
  lengths on the pion mass. The ratios $R_I$'s are the scattering lengths
  normalized to the leading order value.} 
\end{figure}
In Fig.~\ref{fig:a0} the dependence of the two $S$ wave scattering lengths
on the pion mass is shown. The scattering lengths are divided by their
leading order value. The figure shows that large deviations from leading
order occur rather early, particularly for the $I=0$ case. This is an
effect of the particularly strong attractive interaction in the $I=0$
channel, as opposed to the much weaker and repulsive one in the $I=2$
channel. The plot shows that lattice calculations of these quantities
should be made down to rather small values of the quark masses if a
reliable extrapolation to the physical value has to be made.
In comparison, quantities that do not involve two pions in the final state,
like the pion mass and decay constant, show a much milder dependence, see
Fig.~\ref{fig:fpi}. 
There are calculations of the pion scattering lengths already available in
the literature, particularly for the $I=2$ channel which is considerably
easier. Unfortunately they are all done at rather high quark masses. The two
most recent ones are in Ref.~\cite{scattlat}, where reference to earlier
calculations can also be found.

The strong quark mass dependence of the $I=0$ scattering length sounds as a
warning also for future direct calculations of the $K \to \pi \pi$
amplitude. In that case the FSI are even stronger than
for the $\pi \pi$ amplitude at threshold (at $s=M_K^2$ the phase in the
$I=0$ channel is around $40^\circ$), and the quark mass dependence should
presumably be similarly strong -- implying again the need for simulations
at rather small values of quark masses, before a reliable extrapolation to
the physical value can be made.

\subsection{Chiral $SU(3) \times SU(3)$}
\begin{figure}[thb] 
\leavevmode 
\begin{center}
\includegraphics[width=7.2cm,angle=-90]{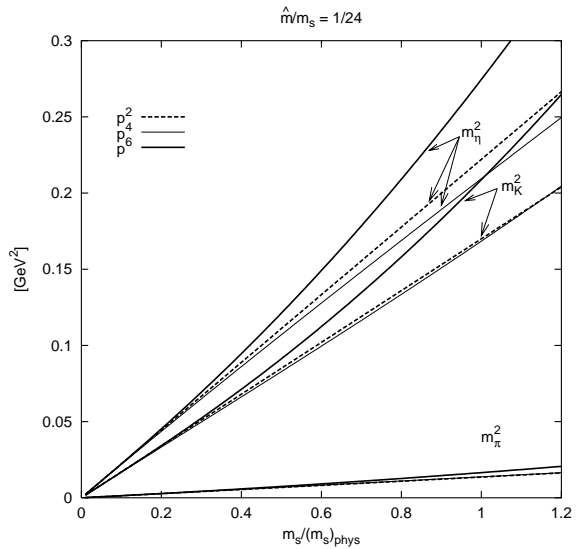}
\end{center}
\vskip -1cm
\caption{\label{fig:msu3}Dependence on the strange quark mass
  of the meson masses, from \protect\cite{ABT}}
\end{figure}

\begin{figure}[thb] 
\begin{center}
\includegraphics[width=7.2cm,angle=-90]{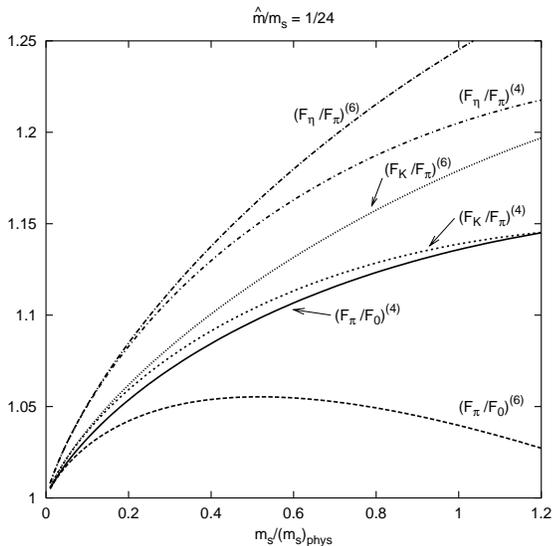}
\end{center}
\vskip -1cm
\caption{\label{fig:fsu3}Dependence on the strange quark mass
  of the meson decay constants, from \protect\cite{ABT}}
\end{figure}

The calculations of quantities at next-to-next-to-leading order in $SU(3)
\times SU(3)$ CHPT are much more complicated, and the convergence of the
series is certainly slower than for chiral $SU(2) \times SU(2)$. Recently,
an analysis of $K_{e4}$ decays at this order has been performed
\cite{ABT}. In this work all the low energy constants of order $p^6$ have
been estimated with resonance saturation, whereas the $O(p^4)$ constants
have been fitted by comparing to the data on $K_{e4}$ form factors. Again,
having fixed all the constants, the authors were also able to consider the
quark mass dependence of various quantities, and particularly of meson
masses and decay constants. These are shown in Fig.~\ref{fig:msu3} and
\ref{fig:fsu3}.
The results of this paper are puzzling as far as the convergence of the
series for masses and decay constants are concerned: the authors find that
the $O(p^6)$ contributions are systematically larger than the $O(p^4)$
ones, as one can see from Figs.~\ref{fig:msu3} and \ref{fig:fsu3}. It is
also surprising how strongly the decay constants depend on the strange
quark mass, including the one of the pions. The reader is referred to the
original paper \cite{ABT} for a discussion of the uncertainties involved in
the determination of the various low energy constants -- as stressed above,
those governing the quark mass dependence of meson masses are the most
difficult ones to control. In view of this, these plots should rather be
viewed as an indication of the qualitative behaviour, rather than precise
quantitative statements. It would of course be very interesting if lattice
calculations could contribute and calculate the dependence on the strange
quark mass of various meson masses and decay constants.

\section{SUMMARY}
In this talk I have reviewed and critically discussed the recent literature
on final state interactions in $K \to \pi \pi$, and a few examples of quark
mass dependence of various quantities as calculated in CHPT at
next-to-next-to-leading order. I believe that the few issues discussed here
are good examples of how important the exchange of information between the
lattice and the chiral communities is at present -- and of how important it
will become in the future.

\section*{ACKNOWLEDGEMENTS}
It is a pleasure to thank M.~B\"uchler, J.~Kambor and F.~Orellana for a
pleasant collaboration on the subject discussed here, and the organizers
for the invitation and the perfect organization of the conference.
The hospitality at the Institute for Nuclear Theory at the University of
Washington (Seattle), where this work was completed, is gratefully
acknowledged.

\end{document}